\documentclass[aps,pra,showpacs,amsmath,amssymb,twocolumn]{revtex4}
\usepackage{epsfig}
\usepackage{graphicx}
\begin{document}

\title{Universality and scaling limit of weakly-bound tetramers}

\author{M. R. Hadizadeh$^1$\footnote{E-mail: hadizade@ift.unesp.br}, 
M. T. Yamashita$^1$, Lauro Tomio$^{1,2}$\footnote{E-mail: tomio@if.unesp.br},
A. Delfino$^2$, T. Frederico$^3$\footnote{E-mail: tobias@ita.br}}
\affiliation{$^1$Instituto de F\'{\i}sica Te\'orica,
Universidade Estadual Paulista, 01140-070, S\~ao Paulo, SP, Brazil\\
$^2$Instituto de F\'{\i}sica, Universidade Federal
Fluminense, 24210-346, Niter\'oi, RJ, Brazil\\
$^3$Instituto Tecnol\'ogico de Aeron\'autica, 12228-900, S\~ao
Jos\'e dos Campos, SP, Brazil}
\date{\today}
\begin{abstract}
The occurrence of a new limit cycle in few-body physics, expressing 
a universal scaling function relating the binding energies of two 
consecutive tetramer states, is revealed, considering a renormalized
zero-range two-body interaction applied to four identical bosons.  
The tetramer energy spectrum is obtained when adding a boson to an 
Efimov bound state with energy $B_3$ in the unitary limit (for zero
two-body binding, or infinite two-body scattering length).
Each excited $N-$th tetramer energy $B_4^{(N)}$ is shown to slide 
along a scaling function as a short-range four-body scale is changed, 
emerging from the 3+1 threshold for a universal ratio 
$B_4^ {(N)}/B_3 \simeq 4.6$, which does not depend on $N$.  
The new scale can also be revealed by  a resonance in the
atom-trimer recombination process. 
%%% 03.65.Ge Solutions of wave equations: bound states  
%%% 21.45.-v Few-body systems
%%% 67.85.Jk other BEC phenomena (three- and four-boson recombinations)
%%% 05.10.Cc Renormalization group methods
\end{abstract}

\maketitle

The rich nature of quantum few-body systems interacting with
short-ranged forces is not shaped only by three-body properties.
Weakly-bound tetramers composed by identical bosons and their
excited states have a characteristic scale, which is independent of
the trimer one, for resonant pairwise interaction in the unitary
limit (zero two-body binding or infinite scattering length
$a\to\pm\infty$). Such property can be revealed by considering the
general case, not constrained by some specific strong short-range
interaction.
The existence of an unsuspected new limit cycle is shown, which is
expressed by an universal function relating the binding energies of
two consecutive tetramer states, $B_4^ {(N)}$ and $B_4^ {(N+1)}$
(where for $N=0$ we have the ground-state) and the corresponding
three-body subsystem binding energy $B_3$ of an Efimov state
\cite{efimov71,efimov09}.
We further derive that the $N+1$ tetramer emerges from the 3+1
threshold for a universal ratio $B_4^ {(N)}/B_3 \simeq 4.6$, which
does not depend on $N$. The tetramers  move as the short-range
four-body scale is changed. The existence of the new scale can be
also revealed by  a resonance in the atom-trimer recombination
process. The resonant behavior arises when a tetramer becomes bound
at the atom-trimer scattering threshold. Furthermore, the independent 
four-body scale implies in a family of Tjon lines~\cite{tjon-line} 
in the general case.

The findings reported here have the purpose to clarify and  advance
the field by recognizing the independent role of a four-body scale
near a Feshbach resonance and its implications for cold atoms. This
issue has been scrutinized in recent theoretical approaches
presented in
Refs.~\cite{meissner04,epl06,hammer_platter,stechernature,incaoprl2009,stecherjpa}.
The relevance of such study is related the experimental
possibilities to explore universal few-body   properties with
tunable two-body interactions. In this respect, it was found experimental 
indications on the existence of two tetramer states associated to an Efimov 
trimer~\cite{ferlaino:140401}, in agreement with theoretical analysis by 
Stecher et al.~\cite{stechernature}.
The studies presented in Ref.~\cite{stechernature} also suggest 
that the only relevant scales near the Feshbach resonance are the two and
three-body ones. 
However, as we are pointing out in the following, this is not enough in order 
to have a full description of the tetramer physics.

Experiments with cold atoms near the Feshbach resonance observed Efimov cycles 
in the position of the resonant three-body recombination as function of the 
scattering length (for recent reports on those findings, see 
Refs.~\cite{phys-ferlaino,efimov09}). Furthermore, it was also found evidences 
for resonant four-body recombination, as discussed in \cite{phys-ferlaino}).
Our results, leading to a new universal four-body limit cycle, show that the 
reported experiments and corresponding calculations have been limited to a 
very specific region of our theoretical predictions, suggesting that new 
physics related to the four-body scale are still unrevealed.

First, one should realize that, in three dimensions, the collapse of
the three-body ground-state, as the two-body range $r_0$ is reduced to
zero (Thomas collapse~\cite{ThPR35}), and the accumulation of three-body 
excited states in the limit that the two-body scattering length $a$ goes 
to infinity (Efimov effect~\cite{efimov71}), which are recognized related 
effects (Thomas-Efimov effect~\cite{th-efimov1988,Lee2007}), 
are manifestations of the sensitivity of the low-energy physics to 
short-range effects, which are parameterized by a three-body scale. 
After such scale has been established, within an appropriate scaling 
approach, as the one proposed in Ref.~\cite{amorim99}, it was shown that 
the dislocation of the three-body scale in respect to the two-body one 
can be clearly revealed by resuming the Efimov plot (left frame of  
Fig.~\ref{fig1}) in a single curve (right frame of Fig.~\ref{fig1}), 
named scaling function ${\cal F}_3^{(N)}$. 
This function  in the zero-range limit depends on $a=\pm 1/\sqrt{B_2}$, 
positive for bound and negative for virtual dimers, in units of 
the natural length scale of a trimer:

\begin{equation}
\sqrt{\frac{B_3^{(N+1)}-\overline{B}_{2}} {B_3^{(N)}}} \equiv {\cal
F}_3^{(N)}\left(\pm\sqrt{\frac{B_{2}}{B_3^{(N)}}}\right),
\label{eq1}\end{equation} where $\overline{B}_2=B_2$ for bound 
(plus sign) two-body system and set to zero for virtual states 
(minus sign).
\begin{widetext}
{\vskip -5cm . \vskip 5cm}
\begin{figure}
\centerline{
\includegraphics[width=12cm]{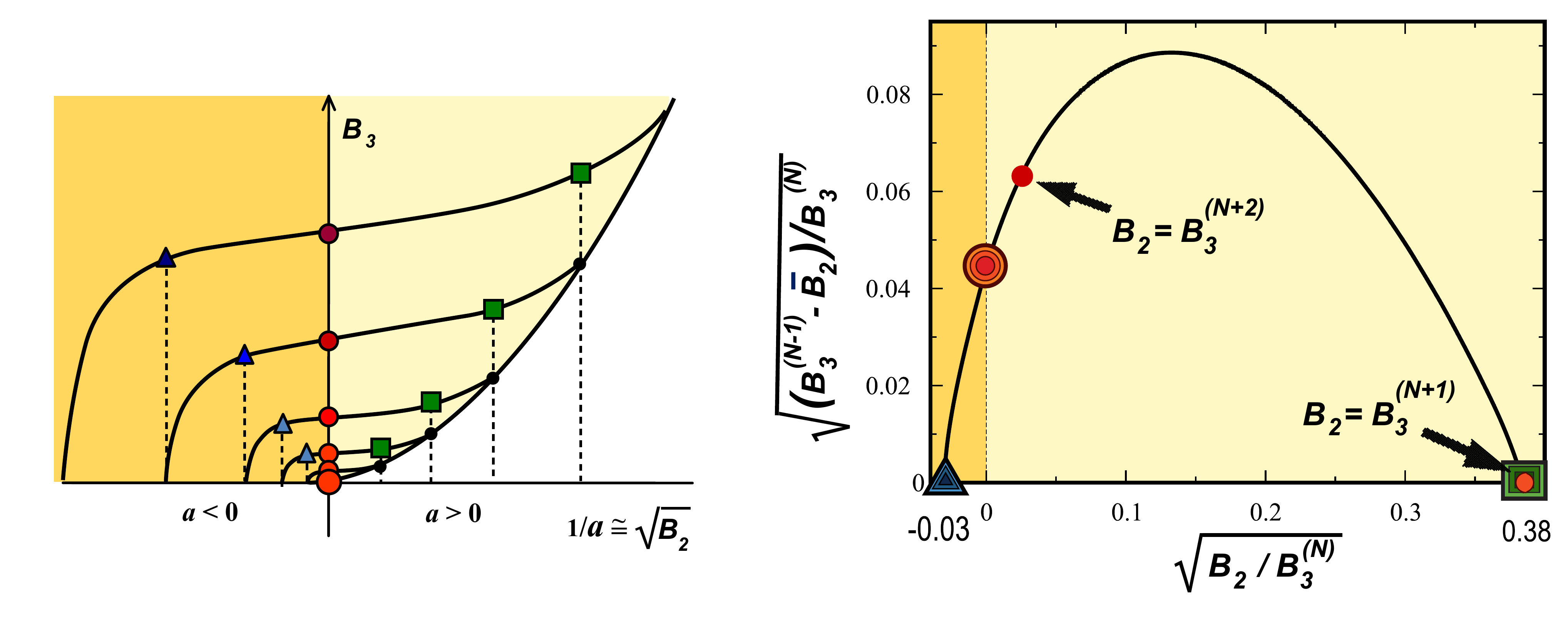}}
\caption{\small{
%(Color on-line) 
Schematic Efimov plot (left) with the corresponding three-body scaling 
function (right), where $B_2 = \hbar^2 /(m a^2)$, with $m$ the atom mass 
(Our units are such that $m=1$ and $\hbar=1$). As shown, the sequence 
of three-body energies $\{ B_3^{(0)}, B_3^{(1)}, B_3^{(2)}, ...\}$ can 
be replaced by just one exact quantitative scaling plot~\cite{amorim99} 
(right). For $B_2=0$, the Efimov tower of weakly-bound $s-$wave states 
of maximum symmetry are geometrically separated by $B_3^{(N+1)} = e^{-\frac{2\pi}{s_0}}B_3^{(N)}$ with $s_0 = 1.00624$ (empty circles).}} 
\label{fig1}
\end{figure}
\begin{figure}
\centerline{
\includegraphics[width=14cm]{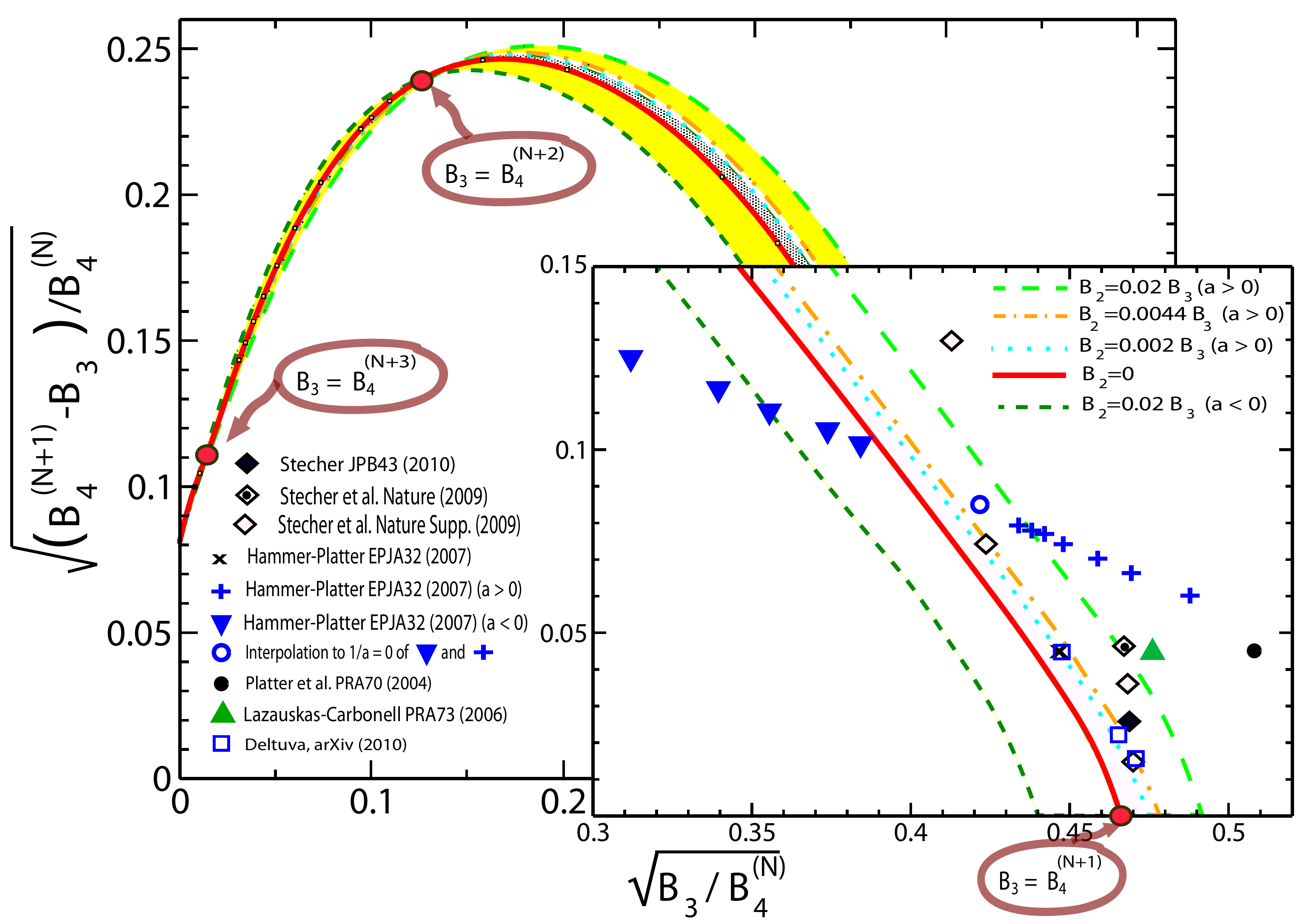}
}
\caption{\small{
Four-body scaling plot for the excited $N+1$ tetramer binding energy, 
$B_4^{(N+1)}$, within the renormalized zero-range model. 
The unitary limit is shown by the solid-red line. The virtual branch
($a<0$) is represented by $B^{virtual}_2/B_3=$0.02 (small-dashed line). 
For non-zero $a>0$, we show three cases: $B_2/B_3=$0.02 
(large-dashed-green line),
$B_2/B_3=$0.0044 (dot-dashed-orange line), and
$B_2/B_3=$0.002 (dotted-blue line). 
The symbols are indicating other model calculations. In the unitary limit,
we show results of Ref.~\cite{hammer_platter} ($\times$), Ref.~\cite{stecherjpa}
(solid diamond), Ref.~\cite{deltuva2010} (empty boxes), and  an interpolation
of the results extracted from Fig. 2 of Ref.~\cite{hammer_platter} 
(empty-blue circle). Near the unitary limit, we show results given in 
Ref.~\cite{stechernature} (dot-black and empty-black diamonds) and in 
Ref.~\cite{lazauskas2006}. 
From this plot, starting from right to left, by decreasing the reference 
three-body energy $B_3$, we show the first cycle starting for 
$B_3=B_4^{(N+1)}$, when $B_4^{(N)}\approx 4.6 B_3$, with a new cycle 
emerging for $B_3=B_4^{(N+2)}$, when $B_4^{(N)}\approx 64 B_3$. }} 
\label{fig2}
\end{figure}
%%%%%%%%%%%%%%%%%%%%%%%%%%%%%%%%%%%%%%%%%%%%%%%%%%%%%%%%%%%%%%%%%%%%
\vspace{-1cm}\end{widetext}
The scaling function turns out to be universal and after few cycles
become independent on $N$; i.e., in the limit $N\to\infty$ it
reaches a renormalization-group invariant limit
cycle~\cite{bira,wilson}. It is one among many possible model
independent correlations between three-body observables for
short-ranged interactions, i.e., when the range $r_0$ goes to zero as
$|a|/r_0\to \infty$. Range effects can become relevant as $|a|/r_0$
decreases (see, e.g. \cite{limit-cycle-rev}).

The left and right frames of Fig.~\ref{fig1} and Eq. (\ref{eq1})
capture the essential physics of the Efimov effect for small but
non vanishing dimer energies, with three-identical particles. 
How this strange and curious picture changes by adding to a weakly 
bound three-boson system another identical boson? Does it exist 
a scaling function, for the ratio between two consecutive four-body 
binding energies $B^{(N+1)}_4/B^{(N)}_4$, in (or near) the limit
$a\to\pm \infty$, having a similar role as the one found for
three-boson systems?  Our answer is yes! For convenience, 
by considering a given three-body energy $B_3$ and 
$a\to\pm \infty$, we will write such a function in analogy with  
 Eq.~(\ref{eq1}), such that it is zero defined for each case 
that the three-body binding energy $B_3$ coincides with one of the 
excited four-body energy states. Therefore, the four-body scaling
function will be defined as
\begin{eqnarray}
\sqrt{\left[\frac{\left(B^{(N+1)}_4 - B_3\right)}{B^{(N)}_4}\right]}=
{\cal F}_4^{(N)}\left(\sqrt{\frac{B_{3}}{{B_4^{(N)}}}}\right) . 
\label{eq2}
\end{eqnarray} 

We found theoretically a simple way to move the four-body 
bound-state energy while keeping the three-body one unchanged. The
essential idea is to recognize that the interacting three-body
subsystem propagation within the four-body system   carries the
three-body scale, i.e., the three-body parameter that drives the
Thomas-Efimov effect. Such parameter fixes the three-body
properties. The four-body scale parameterizes the physics carried by
the ultraviolet momentum region of the four-boson propagation
between different fully interacting three-body cluster propagation
or disjoint two-body clusters propagation in the Faddeev-Yakubovsky
(FY) equations (see technical details in the appendix). This is the
essence of the renormalization procedure that have been developed in
a previous work of our group~\cite{epl06}.

We explore the universal physics attached to the new parameter, by
solving the four-body bound-state equations, while preserving the
three-body binding energies and corresponding scale unchanged, in
the unitary limit. Our conclusions are supported by  extensive and
accurate numerical calculations of the ground and excited tetramer
states energies and momentum space wave functions. We verified the
sensitivity of the four-body bound-state energy to the four-body
scale by moving it. We found that excited four-body states come out
from the atom-trimer scattering threshold as the four-body parameter
is driven to short distances or to the ultraviolet momentum region.
Figure~\ref{fig2} illustrates this phenomenon. We see a close analogy to the
left frame of Fig.~\ref{fig1} (for $a>0$), where the three-body excited states
comes from the two-body scattering threshold as the dimer binding
decreases $|B_2|\to 0$. The  binding energies of two consecutive
tetramers resume themselves into a single curve depicted in 
Fig.~\ref{fig2}. We observe that the convergence with $N$ is fast towards the
limit cycle, namely the new four-boson scaling function.

The universal property of the new scaling function can be
appreciated by comparing it to different theoretical approaches in
the unitary limit as shown in Fig.~\ref{fig2}. The theoretical frameworks
are very much distinct from ours. Stecher et al performed
calculations with different model potentials using adiabatic
hyperspherical approximation and correlated Gaussian basis set
expansion~\cite{stechernature,stecherjpa}. Hammer and Platter used
one-term s-wave separable potentials within effective field theory
where a repulsive three-body force stabilizes the trimer energy
against collapse~\cite{hammer_platter}, and recently Deltuva
also calculated the position of the four-body resonances with one- and
two-term separable potentials~\cite{deltuva2010}. Irrespectively, to which
trimer the tetramers are nearest below, the energies scale according
to the scaling plot shown in Fig.~\ref{fig2}.

Actually, we are confirming the existence of an independent four-body 
short-range parameter as suggested in \cite{epl06}, with new results 
exploring the excited tetramer spectrum, with precise and accurate 
numerical results. { The new scale appears not only for the relation
between the first two lower tetramer states but also for the 
relation between other consecutive more excited states.} 
The available results of different calculations and
theoretical methods, which are close to the unitary limit, confirm
the universal scaling shown in Fig.~\ref{fig2}. Besides the fact that such
results are limited to a small region of our plot, the existence of
an independent four-body scale is clearly catch. It shows the
equivalence of our results with the ones found with other methods.
Our theory distinguishes itself by the straightforward ability to
keep fixed the three-body energies while the four-body parameter is
shifted. With potential models, one should dial the two- and three-body
forces maintaining the three-body states unaltered as well the
two-body properties and check for the change in the binding of the
tetramer states. A short-range four-body force will do the job, with
fixed two- and three-body properties. Indeed, as we show through the
scaling plot in Fig.~\ref{fig2}, other theoretical approaches have embedded
the four-body scale.

Calculations that have been done within other different potential
models~\cite{meissner04,hammer_platter,wang,stechernature} should
allow a wider exploration of the ratios between three and four-body
energies, in order to observe effects from the four-body independent
scale. 
{However, in the parametrization of more realistic interactions, 
it is obvious not so straight to 
consider a fixed three-body energy, and obtain the corresponding 
four-body spectrum near the unitary limit}.

{\it What does the literature say about four-body systems and Efimov
effect with short-ranged interactions?}

Let us remind the past discussions on the {\it true} Efimov effect
and universality or scaling in four-body systems. The pioneering
work of Amado and Greenwood~\cite{amado} have addressed the first
issue and it was followed by the works of Fonseca and
Adhikari~\cite{fonseca-adh} and Tjon and Naus~\cite{tjon1}, both
performed within the Born-Oppenheimer approximation for three
massive bosons and a light one. Amado and Greenwood estimated the
trace of the kernel of the four-body integral equation in momentum
space and showed that there is no infrared divergence from that they
concluded against the true Efimov effect. However, the momentum
integrals should have implicitly an ultraviolet cutoff (the
four-body one) to regulate it. The trace diverges as the cutoff runs
to infinity which does not conflict with the existence of an
infinite number of four-boson bound states. Tjon addressed the
second issue by showing that the alpha-particle and the triton
binding energies are linearly correlated for different short-ranged
two-nucleon potentials in a model independent way~\cite{tjon-line}.

Conclusions drawn within the nuclear physics context are not general
enough to assure that four-boson systems with short-ranged
interactions do not have a new independent scale, as models for the
nuclear interaction are pretty much constrained. Also the
possibility of experiments with few-nucleon systems to explore
general aspects of few-body physics due to a short-range scale are
limited by the strong nuclear repulsion of the potential core. For
example, it is well known that the three-body Efimov prediction on
the existence of a series of low-energy three-body states, for zero
two-body binding, was never undistinguished recognized for
three-body nuclear systems. The most promising exploration in this
aspect has been connected with the discoveries of exotic halo-nuclei
systems. Finally, the Efimov effect is being recognized as a real
physical phenomenon, in view of the recent cold-atom experiments
performed with two-body scattering lengths varying by many decades.

Our results imply in the existence of a family of Tjon lines~\cite{tjon-line} 
with slopes determined by the new scale. The separation between
consecutive states also depends on the four-body parameter. At the
threshold to bind the excited $N+1$ state, we get
$B^{(N)}_4\simeq4.6 B_3 \ ,$ solution of ${\cal
F}_4^{(N)}\left(\sqrt{B_{3}/B_4^{(N)}}\right)=0$, which agrees, as
the tetramer scaling plot in Fig.~\ref{fig2} shows, with the existing
calculations at the unitary limit.

{\it What are the effects of a four-body scale in cold atoms, how to
observe them?}

The existence of a four-body scale has an impact on cold atom
physics where low-energy and universal properties of few-body
systems are intensively explored, i.e., in systems with
characteristic length scales much larger than the interaction range.
Universal properties were in fact observed with trapped cold-atoms
near a Feshbach resonance, by dialing the atom-atom scattering
length $a$ over several order of magnitude. From their resonant
contribution to inelastic collisions and the corresponding trap
losses, the experiments have in fact confirmed both, the presence of
geometrically separated Efimov trimers (see e.g. \cite{Kraemer}),
and two associated
tetramers~\cite{ferlaino:140401,Zaccanti,hulet2009}. The four-body
recombination experiments with tunable interaction, which are in
agreement with some theoretical predictions~\cite{stechernature},
hopefully can also explore regions, for large $|a|$, where a
three-body binding is much smaller than the four-body one, in order
to verify the four-boson scaling behavior we are predicting in
Fig.~\ref{fig2}. Four-body observables, like four-boson recombination rates
or atom-trimer or dimer-dimer scattering lengths, can exhibit
correlations not constrained by  one low-energy s-wave three-boson
observable and $a$, as exemplified in Fig.~\ref{fig2}, by the limit cycle
for two consecutive tetramer states.

The coupled channel nature of  the Feshbach resonance induces, by
the reduction of the problem to a single channel, three and
four-atom potentials, which can drive independently the
corresponding physical scales.  The induced two-body interaction in
the open channel is attractive, producing a near-threshold S-matrix
pole corresponding to a weakly-bound or virtual two-body state. The
induced three-body interaction can be either attractive or
repulsive, because it produces three-body amplitudes where the
resonance is shared by different two-particle particle subsystems.
For example, the interaction excites one atom of a pair in the
closed channel that interacts with the third atom in the closed
channel of a new pair. The magnitude of the few-atom force should
increase by approaching the Feshbach resonance due to vanishing
energy denominators of the virtual intermediate propagations of the
subsystems in the closed channel~\cite{epl06}. As one cannot a
priori exclude the relevance of three- and four-atom effective
interactions in the open channel, one could argue why the observed
positions of the four-atom resonances seem to agree with the
universal theory, with no need of a new scale beyond the three-boson
one~\cite{phys-ferlaino,hulet2009}. We suspect that those
experimental results are in a region where only two-body potentials
are important, with much smaller or slowly varying three-body
interactions.

However a signature of many-body forces near the Feshbach resonance
is emerging. The need for a dislocation of the three-body parameter
was found in the recombination rates of $^{39}$K by Zaccanti et al.
\cite{Zaccanti} and in $^7$Li by Pollack et al. \cite{hulet2009}
when crossing the Feshbach resonance, also in an experiment of
atom-dimer loss in an ultracold trapped gas of a mixture $^6$Li
atoms in three hyperfine states performed by Nakajima et al.
\cite{nakajimaprl10}. These works found that the locations of the
recombination peaks disagree with the predictions of the universal
theory with a fixed value of the three-body parameter. In the case
of $^6$Li even contributions from two-body nonuniversal properties
were excluded, and more, it was shown that different two-body models
lead to a model independent interpretation of the nonuniversal
physics of the Efimov trimers. Therefore, there are striking
evidences for a nonuniversal nature of the short-range three-body
physics through the variation of the three-body parameter in
coldatom experiments. These experiments led us to convey that: {\it
i)} nonuniversal short-range physics beyond two-body properties
emerges near a Feshbach resonance; and {\it ii)} model independent
predictions are still possible under the situation {\it i)}. This is
an indication of the presence of a short-range three-body force,
that moves the three-body scale. Correlations between observables
survives the change in the three-body parameter supporting the
conclusion of {\it model independence} found by Nakajima et al.

Once a short range three-body force appears the very same mechanism
produces effective four-body forces acting in the open channel. In
the presence of three- or even four-body forces the three and
four-body short range scales can move and detach the physics of
trimers from tetramers. Therefore makes sense to search for effects
that are tied to the different scales that parameterize the short
range part of effective interaction in the open channel and are
driven by different forces in trapped cold atoms near a Feshbach
resonance.

In order to verify experimentally the scaling due to the four-body
parameter, we suggest to tune the large and negative atom-atom
scattering lengths in a region where Borromean trimers are possible
without the formation of weakly-bound dimers. In this case trap
losses due to inelastic dimer-dimer and atom-dimer collisions are
absent.  A resonant contribution to trap losses due to atom-trimer
inelastic collisions occurs when the tetramer goes to the
atom-trimer scattering threshold, i.e., its binding energy becomes
equal to the trimer one. The position of the atom-trimer resonance
is  not only a function of the atom-atom scattering and the
three-body scale, but it will also depends on the new four-body
scale. Therefore, in this case the Efimov ratio of 22.7 between the
values of the scattering length, corresponding to the position of
consecutive resonances, is not assured. The ratio between the
scattering lengths where two consecutive tetramers becomes unbound
depends not only on the three-body parameter but also on the
four-body one. The same is true for the ratio between the scattering
lengths where the Borromean trimer and tetramer disappears.

The physics of four-atom systems  close to a Feshbach resonance
demands one three and one four-body scale which move as the large
scattering length is tuned,  driven by different few-body forces. As
odd it can sound, universality is still shaping the physical
quantities:  the four-body short-range scale resumes itself in new
limit cycles, which brings universal properties through scaling
functions by means of correlations between two different low-energy
observables of weakly-bound tetramers, as the binding
energies of two consecutive tetramers. A strong experimental
evidence revealing the new physics will be the observation of
resonant inelastic collisions in the atom-trimer channel with large
and negative scattering lengths.

\section*{Acknowledgements}
We thank Hans-Werner Hammer, Randy Hulet and Lucas Platter for
helpful information.
We also acknowledge partial financial support
from the Brazilian agencies Funda\c c\~ao de Amparo \`a Pesquisa do
Estado de S\~ao Paulo and Conselho Nacional de Desenvolvimento
Cient\'\i fico e Tecnol\'ogico.

\appendix\section{Renormalized Faddeev-Yakubovsky equations}

We solve the four-boson Faddeev-Yakubovsky equations in momentum
space, for a zero-range potential by considering a regularizing
procedure, which is based in a subtraction approach where it is introduced
a renormalizing momentum scale $\mu_{4}$, such that the four-body free
Green's function $G_0(E)$ is replaced by $G_0(E)-G_0(-\mu_{4}^2)$.
This approach, detailed in Ref.~\cite{epl06}, is generalizing
the subtracted equation for trimers, given in Ref.~\cite{AdPRL95}:
\begin{multline} 
 \tau^{-1}(e^{(1,3)}_{2} )|{\cal K }_{ij,k}^l\rangle-
 {\cal G}^{(3)}_{ij;ik} |{\cal K} _{ik,j}^l\rangle
-  {\cal G}^{(3)}_{ij;jk}|{\cal K} _{jk,i}^l\rangle =\\ =
  {\cal G}^{(4)}_{ij;ik}\big[ |{\cal K}_{ik,l}^j\rangle +
|{\cal H}_{ik,jl}\rangle\big]+  {\cal G}^{(4)}_{ij;jk} \big[|{\cal
K}_{jk,l}^i\rangle + |{\cal H}_{jk,il}\rangle\big],
\\
 \tau^{-1}(e^{(2,2)}_{2}) |{\cal H} _{ij,kl}\rangle=  {\cal
G}^{(4)}_{ij;kl}\big[ |{\cal K }_{kl,i}^j\rangle + |{\cal
K}_{kl,j}^i\rangle +|{\cal H}_{kl,ij}\rangle\big] , \label{subs2}
\end{multline}
where $e^{(1,3)}_{2}=E-E_{ij,k}-E_l$ and
$e^{(2,2)}_{2}=E-E_{ij,kl}-E_{kl}$ are
 the two-body subsystem energies in the 3+1 and 2+2 partitions,
 respectively. The projected Green's function operators are
$ {\cal
G}^{(N)}_{ij;ik}:=\langle\chi_{ij}|{G}^{(N)}_{0}|\chi_{ik}\rangle $
with $N$ equal 3 or 4, with the subtracted Green's functions given
by $G^{(3)}_0=[E-H_0]^{-1}-[-\mu_{3}^2- H_0]^{-1}$ and
$G^{(4)}_0=[E- H_0]^{-1}-[-\mu_{4}^2- H_0]^{-1}$. The two-boson
scattering amplitude with a proper normalization is given by
$\tau^{-1}(x)=2\pi^{2}\left[a^{-1}-\sqrt{-x}\right]\to
-2\pi^{2}\sqrt{-x}$ in the unitary limit and $\langle
\vec{q}_{ij}|\chi_{ij}\rangle=1$. The four-boson integral equations
for the reduced FY amplitudes are projected to states of total
angular momentum zero. The corresponding set of homogeneous integral
equations, which after discretization turns into a huge matrix
eigenvalue equation, is solved by a Lanczos-like method, 
by iteration. This method is very efficient for few-body problems 
(see \cite{hadi2007}). For the discretization of continuous momentum 
and angle variables we have used Gaussian-quadrature grid points 
with hyperbolic and linear mappings, respectively.

{\small
\begin{table}
\label{tab1}
\caption{Tetramer ground and excited state binding energies for
different four-body scales, considering the unitary limit
(infinite two-body scattering length).
}
\begin{center}
\begin{tabular}{lccccccccccccccl}
\hline\hline \centering
 && $\mu_{4}/\mu_{3}$  &&   $B_4^{(0)}/B_3$ &&   $B_4^{(1)}/B_3-1$ &&  $B_4^{(2)}/B_3-1$ \\
\hline 
 &&   1 && 3.10  \\
 && 1.6 && 4.70   && 7.1$\times 10^{-4}$ \\
 &&   5 && 12.5   && 0.531 \\
 &&  10 && 24.6   && 1.44 \\
 &&  21 && 63.5   && 3.62 && 3.2$\times 10^{-4}$&&  \\
 &&  40 && 184    && 7.65 && 0.203 \\
 &&  70 && 5.20$\times 10^{2}$   && 12.9 && 0.629 \\
 && 100 && 1.04$\times 10^{3}$   && 20.5 && 1.17 \\
 && 200 && 4.06$\times 10^{3}$   && 50.8 && 2.86 \\
 && 300 && 9.11$\times 10^{3}$   && 101 && 4.75 \\
 && 400 && 1.62$\times 10^{4}$   && 153 && 6.28 \\
\hline\hline
\end{tabular}
\end{center}
\end{table}
}

The two-, three- and four-body momentum scales 
are $a^{-1}$, $\mu_{3}$ and $\mu_{4}$, respectively, which give us the ground and 
excited tetramer binding energies, for $a=\pm\infty$ depending on the momentum 
scales as $B_4~= \mu_{4}^2{\cal {B}}_4\left(\mu_{3}/\mu_{4}\right)$. 
The four-body scaling function, given by Eq.(\ref{eq2}), is obtained when 
$\mu_{3}$ and $\mu_{4}$ are replaced by $B_3$ and $B_{4}^{(N)}$. 
By keeping fixed the two- and the three-body scales and moving independently 
the four-body scale, we obtain the tetramer binding energies for the ground 
and excited states.

In Table I, we have listed our numerical results for tetramer ground 
and excited state binding energies, considering the unitary limit
(infinite two-body scattering length) for scale ratios 
${\mu_{4}}/{\mu_{3}}$ varying from 1 to $400$. The results correspond
to the solid-red line shown in Fig.~\ref{fig2}.
According to the results, 
${\mu_{4}}/{\mu_{3}}\cong 1.6$ is the threshold for the first tetramer 
excited state, and ${\mu_{4}}/{\mu_{3}} \cong 21$ the threshold for 
the second tetramer excited state.  The third tetramer excited state 
should emerge close to ${\mu_{4}}/{\mu_{3}} \approx 240$. 
The binding energy ratio of two consecutive tetramers at these
critical values are ${B_4^{(N)}}/{B_4^{(N+1)}}\cong4.6$, which
is consistent with the results obtained by other authors. 

By dropping the H-channel or cutting the momentum integration below
$\mu_4$, the dependence on the new scale for the four-boson physics
will be minimized (may be even completely removed!). Therefore,
unreasonable selection of the cut-off values in mapping of momentum
variables and for very large $\mu_4$  can lead to convergence in the
four-body binding energies rather than collapse. So, for mapping of
momentum variables one should not only consider large enough cut-off
values, consistent with used four-body scale to achieve cut-off
independent results, one should also consider reasonable number of
mesh points close to zero momentum region. Since the iteration of
coupled equations (\ref{subs2}) requires a very large number of
multi-dimensional interpolation Yakubovsky components, we have
used the Cubic Hermite Splines for its accuracy and high
computational speed.

\end{document}